\documentclass[journal,comsoc]{IEEEtran}
\usepackage[utf8x]{inputenc}
\usepackage[spanish, english]{babel}

\usepackage{subcaption}

\usepackage{url}

\usepackage{graphics,graphicx} %paquetes grÃ¯Â¿Â½ficos estÃ¯Â¿Â½ndar%
\usepackage{wrapfig} %paquete para grÃ¯Â¿Â½fica lateral%
\usepackage[rflt]{floatflt} %figuras flotantes%
\usepackage{graphpap}	%comando \graphpaper en el entorno picture%
\usepackage{graphicx}

\usepackage{tabularx} % in the preamble
\usepackage{multirow}

\usepackage{textcomp}
\usepackage{listings}
\usepackage{xcolor}

\colorlet{punct}{red!60!black}
\definecolor{background}{HTML}{FDFDFD}
\definecolor{delim}{RGB}{20,105,176}
\colorlet{numb}{magenta!60!black}

\lstdefinelanguage{json}{
	basicstyle=\normalfont\ttfamily,
	numbers=left,
	numberstyle=\scriptsize,
	stepnumber=1,
	numbersep=8pt,
	showstringspaces=false,
	breaklines=true,
	frame=single,
	backgroundcolor=\color{background},
	literate=
	*{0}{{{\color{numb}0}}}{1}
	{1}{{{\color{numb}1}}}{1}
	{2}{{{\color{numb}2}}}{1}
	{3}{{{\color{numb}3}}}{1}
	{4}{{{\color{numb}4}}}{1}
	{5}{{{\color{numb}5}}}{1}
	{6}{{{\color{numb}6}}}{1}
	{7}{{{\color{numb}7}}}{1}
	{8}{{{\color{numb}8}}}{1}
	{9}{{{\color{numb}9}}}{1}
	{:}{{{\color{punct}{:}}}}{1}
	{,}{{{\color{punct}{,}}}}{1}
	{\{}{{{\color{delim}{\{}}}}{1}
	{\}}{{{\color{delim}{\}}}}}{1}
	{[}{{{\color{delim}{[}}}}{1}
	{]}{{{\color{delim}{]}}}}{1},
}

\usepackage{adjustbox}
\usepackage{longtable}

\usepackage[spanish]{babel}

\usepackage[english, plain]{fancyref} % plain, 
\fancyrefchangeprefix{\fancyreftablabelprefix}{table}
\newcommand*{\fancyrefpartlabelprefix}{part}
\fancyrefchangeprefix{\fancyrefpartlabelprefix}{part}

\usepackage{enumitem}

\newcolumntype{P}[1]{>{\RaggedRight\hspace{0pt}}p{#1}}
\newcolumntype{M}[1]{>{\RaggedRight\hspace{0pt}}m{#1}}

\newcolumntype{Q}[1]{>{\centering\let\newline\\\arraybackslash\hspace{0pt}}p{#1}}
\newcolumntype{N}[1]{>{\centering\let\newline\\\arraybackslash\hspace{0pt}}m{#1}}

\newcolumntype{R}[1]{>{\RaggedLeft\hspace{0pt}}p{#1}}
\newcolumntype{O}[1]{>{\RaggedLeft\hspace{0pt}}m{#1}}

\usepackage{amsmath}
\usepackage[cmintegrals]{newtxmath}

\usepackage{subcaption}
\usepackage{graphics,graphicx} %paquetes grï¿½ficos estï¿½ndar%
\usepackage{float}
\usepackage{tabularx}

\usepackage{bm}
\usepackage{textpos}

\usepackage{hyperref}
\hypersetup{
    colorlinks=true,
    citecolor=black,
    linkcolor=black,     
    urlcolor=cyan,
}

   \author{
		\IEEEauthorblockN{Carlos Baena, Sergio Fortes, Eduardo Baena, Raquel Barco}

		\fontsize{8pt}{10pt}\selectfont
		
		\IEEEauthorblockA{%\IEEEauthorrefmark{1}
			Universidad de Málaga, Andalucía Tech, Departamento de Ingeniería de Comunicaciones,\\ Campus de Teatinos s/n, 29071 Málaga, España
			\\\{jcbg, sfr, ebm, rbm\}@ic.uma.es}
		
		\vspace{-2\baselineskip} 
}

\begin{document}

\title{{Estimation of Video Streaming KQIs for Radio Access Negotiation in Network Slicing Scenarios}}

% Submission guidelines
% https://www.comsoc.org/publications/journals/ieee-comml/ieee-communications-letters-submit-manuscript  

\markboth{First submitted to ``IEEE Communication Letters" in September, 2019}%
{}

% Location-Aware Communications for 5G Networks: How location information can improve scalability, latency, and robustness of 5G,?

\maketitle 

% \begin{textblock}{hhsizei}(hhposi,hvposi)

\begin{textblock*}{\textwidth}(0cm, -5.2cm)
\small {\color{red}
 This is the author’s version of an article that has been accepted for publication in IEEE Commun. Letters. Changes were made to this version by the publisher prior to publication. The final published article is available at \url{https://doi.org/10.1109/LCOMM.2020.2979713} }
\end{textblock*}

\begin{textblock*}{\textwidth}(0cm, 20 cm)
\small {\color{red}\textcopyright 2020 IEEE.  Personal use of this material is permitted.  Permission from IEEE must be obtained for all other uses, in any current or future media, including reprinting/republishing this material for advertising or promotional purposes, creating new collective works, for resale or redistribution to servers or lists, or reuse of any copyrighted component of this work in other works.}
 
\end{textblock*}

\begin{abstract}
%The use of multimedia content has hugely increased in recent times, becoming in one of the most used and important service for the users of mobile networks. Consequently, network operators struggle to optimize their infrastructure to support the best video service-provision to the end user. As an additional dimension, 5G introduces the concept of network slicing as a new paradigm that presents a completely different view of the configuration and optimization of the network. In this, ``slices", this means, specific sets of resources allocated for certain type of users and services, %are agreed between the infrastructure operator and the ``verticals" (direct providers of service to end-users), based on different target end-user requirements. A main challenge of this scheme is to establish which specific resources can provide the agreed quality of service. To do so, the present article presents a system for the estimation of Video Streaming Key Quality Indicators (KQIs) based on network low-layer configuration and metrics.  

The use of multimedia content has hugely increased
in recent times, becoming one of the most important services for the users of mobile networks. Consequently, network operators struggle to optimize their infrastructure to support the best video service-provision. As an additional
challenge, 5G introduces the concept of network slicing as a new  paradigm that presents a completely different view of the network configuration and optimization.
A main challenge of this scheme is  to establish which specific resources would provide the necessary quality of service for the users using the slice. To address this, the present work presents a complete framework for this support of the slice negotiation process through the estimation of the provided Video Streaming
Key Quality Indicators (KQIs), which are calculated from network low-layer configuration parameters and metrics. The proposed estimator is then evaluated in a real cellular scenario.

\end{abstract}

 
    %\vspace{-\baselineskip} 
    
    	\section{Introduction}
%	The different needs of society over the time have been reflected in mobile network evolution. Some of them, such as making a call from anywhere or having broadband internet access, have been covered by the different generations of mobile network. However, nowadays, new services as video on demand are emerging, greatly promoted by platform such as Youtube, Netflix o Twitch, which drives the upgrade of operational networks.
    
	%These video platforms use the adaptive video protocol, which is known as DASH (Dynamic Adaptive Streaming over HTTP)\cite{DASH}, whose operation unlike the progressive download, allows a dynamic bitrate selection by the software client. This is possible thanks to the design of this technology, where the video is coded in different qualities and splitted in segments. In this way, client asks periodically different video segments of one or other quality, based on network capabilities and available buffer in each moment. %Moreover, DASH presents good compatibility with many internet infrastructure by the use of HTTP (HyperText Transport Protocol).
	
	%With the arrival of the fifth generation mobile networks, better known as 5G, it is followed the creation of a flexible network which improve these type of services, although it is proposed three uses cases types depends on requirements:  enhanced Mobile Broadband (eMBB), massive Machine Type Communications (mMTC) and Ultra Reliability and Low Latency Communications (URLLC).
	Fifth generation mobile networks (5G) are expected to allow very flexible network configurations, able to provide connectivity to different services with heterogeneous requirements in an optimal way. Here, three main service categories are expected to be the main target for 5G provision: enhanced Mobile Broadband (eMBB), massive Machine Type Communications (mMTC) and Ultra Reliability and Low Latency Communications (URLLC).

	5G does not intend to offer these different services over an unique radio interface and set of resources, but rather to allocate and configure different resources in order to fulfill their differentiated requirements. From this, the ``network slicing'' concept arises, which consists in a virtual division and sharing of the network elements \cite{netSlicing}.
	
	In this way, the slicing of the network facilitates network operators to offer resources fitted to specifics vertical industries, generally referred simply as ``verticals'' \cite{Vertical}. These verticals (e.g. a car manufacturer, a venue administrator, a shopping mall management, a factory owner, etc.) aim to agree with the operator about a specific quality of service to be provided to its associated end users (e.g. a specific set of vehicles, the attendees to a sport match in a stadium, the customers in a mall, the robots and sensors in a factory, etc.).
	
	In this scheme, the classical approach where cellular networks were monitored based on performance metrics coming from low layers (e.g. physical, link) of the protocol stack focusing on their radio access indicators (e.g. MAC throughput, radio quality, etc.) become insufficient to provide a proper view of how the network is going to support the end-to-end (E2E) requirements of the verticals. In fact, the process of slice negotiation of resources between them and the operators is expected to be focused on E2E service-specific metrics (e.g. video resolution). Such as application-layer metrics are known as key quality indicators (KQIs)\cite{morel2017quality}. In this new approach, management task becomes more complex due to the challenge of obtaining KQIs during network operation.
	
	%This new concept, network slicing, allows to obtain a clearer vision of the trend that operators are following in order to set up and optimize futures cellular networks. Traditionally, these networks have been optimized based on performance indicators (Key Performance Indicators, KPIs), which are combinations of classic performance counters. These indicators are generally measured in low layer of the protocols stack (link, network...), with what they do not let to know with detail the optimum quality of experience perceived by the service user. The new paradigm based in network slicing of the 5G networks allow to optimize the quality of experience end-to-end (E2E) by the definition and analysis of quality indicators (Key Quality Indicators, KQIs) for each service \cite{morel2017quality}.

	The wide adoption of encryption in protocols of high layers, as well as the limited and generally not available access to the application logs in the user equipment (UE) make the full acquisition of KQIs during network operation unfeasible. As a result of this, the adoption of tools able to estimate KQIs based on metrics and configuration parameters coming from lower layers of the cellular network is deemed necessary. These are available to the operator through the control and management planes. Such an approach will allow network management actions aiming at improving the E2E performance of specific services as well as the proper assignment of resources and maintenance of the slices negotiated between verticals and operators.

	In this scope, there have been some approaches for video streaming services. In \cite{pan2018qoe}, a system able to estimate the bitrate of Youtube encrypted video streaming over HTTPS is proposed. In works such as \cite{jia2016measuring}, \cite{chen2014method} or \cite{de2014qoe}, models to predict the Quality of Experience (QoE) of this service are presented, using KPI and HAS (HTTP Adaptive Streaming) profiles. However, all those references are focused on QoE unique scores per service, and not on multiple KQIs, which are the ones expected to be used in network slicing as a base to deal with a slice agreement as they provide a far better granularity of the service provision. In \cite{vaser2015qos} the qualitative relationship between KPIs and KQIs for video streaming and voice services is analyzed, not providing however tools for their numerical estimation and focusing instead on QoE expressions in non-slicing scenarios.

Therefore, up to the authors knowledge, no previous works have addressed the challenge of translating application-layer requirements to specific radio configuration and resources in slicing scenarios. Beyond this state of the art, the present work proposes a novel framework for the support of the negotiation, establishment and maintenance of network slices based on the estimation of the KQIs of video streaming via regression models. The proposed estimation system is then evaluated in a real cellular network. In this way, section \ref{sec:System} presents the general architecture of the proposed framework, detailing its elements. In section \ref{sec:evaluation} the tasks of estimating the KQIs from lower-layer metrics and configuration management parameters (CMs) are defined and evaluated for different machine learning (ML) modeling techniques and using in a real cellular network testbed. Finally, section \ref{sec:conclusion} presents the conclusions of the work.

	\section{Proposed system}
	\label{sec:System}
    \vspace{-0.3\baselineskip} 
	%In this section it is presented the proposed system which allows to configure network by the estimation of KQIs based on measured KPIs. Specially, the system is focusing in the video streaming service.
	%The proposed system, which it is illustrated in figure \ref{fig:arq},it is presented as closed loop where it can be differentiated 5 different blocks: network configurator, indicator extraction tool, modelling tool and graphic user interface.
    %Figure \ref{fig:arq} presents the proposed KQI estimation framework for the support of network slice negotiation. This is composed of four main different blocks: service experience acquisition (SEA), Modelling System (ModSys), Operator Slice Negotiation Agent (OSNA) and the Dynamic Slice Allocation (DySA). 
    %Their logical functionality as well as details on their implementation in a real high-dense indoor cellular network (composed of 12 picocells and the network core \cite{smartCampus}) is provided.

    As described in \cite{D32} slice negotiation processes start when a vertical industry agent demands a set of E2E service-oriented requirements in terms of KQIs, e.g. 90\% of the time at 1440p resolution for the user of the slice \cite{D32}. The price of the associated slices has to be set by the operator. Both vertical and operator then start an iterative process of negotiation where the vertical might reduce the requirements and/or duration of its demand and therefore the operator would readjust its price. After one or several iterations, performed by their automated agents, a final price and set of E2E KQIs should be agreed on. After that, the operator will establish the slice with the set of radio and core resources necessary to support the agreed KQIs.
          \vspace{-0.8\baselineskip} 
        \begin{figure}[h] %[btp]
		\centering
		\includegraphics[width=0.9\columnwidth]{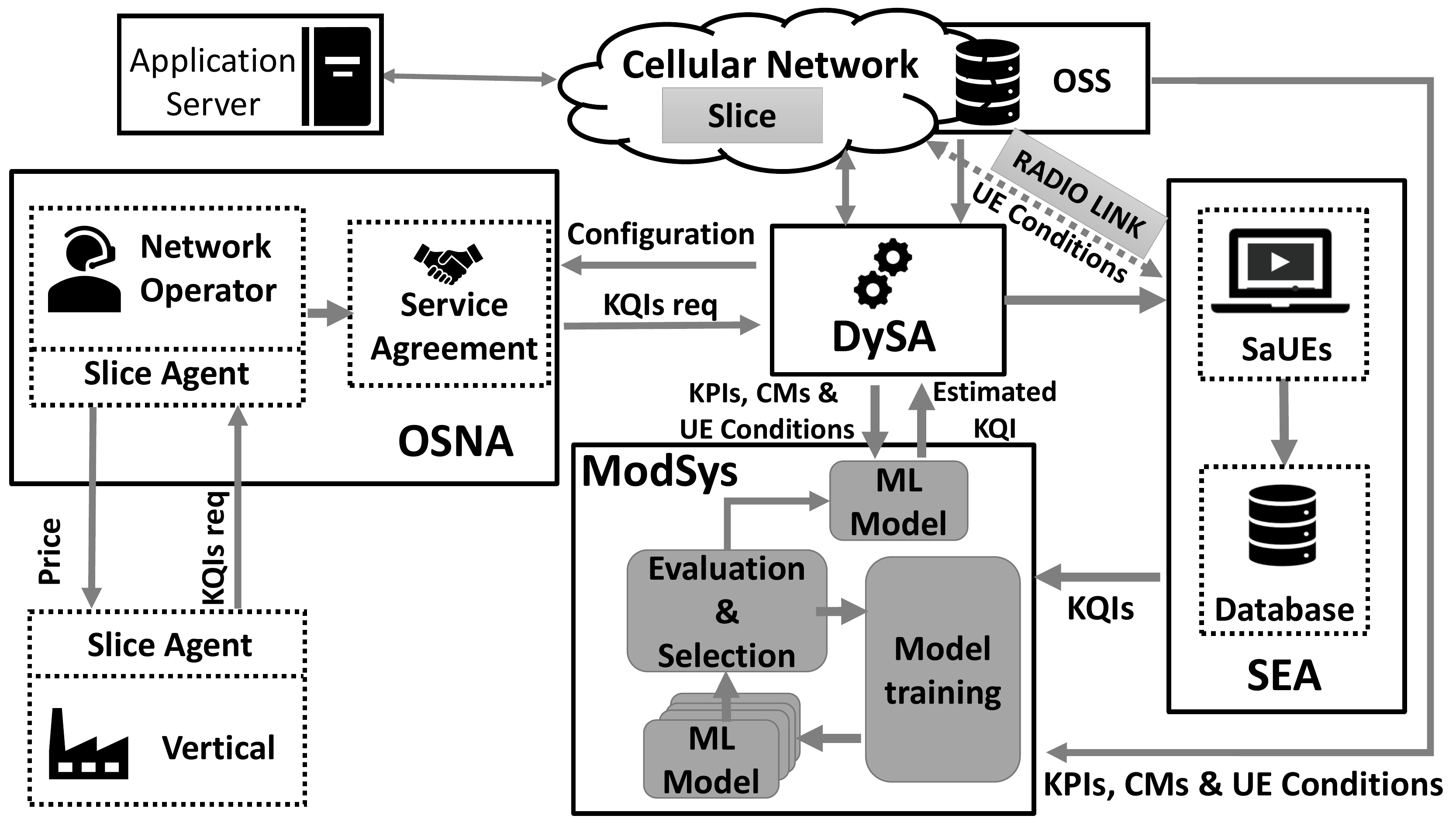}
		\caption{System architecture}
		\label{fig:arq}
	\end{figure}
    
    During the negotiation process as well as for the initial configuration of the slice and its maintenance, the operator must be able to estimate the KQIs that the UEs will experience for certain configurations of the network in their available radio conditions. 
    
    Where previous approaches assumed a small set of fixed and known configuration options, the objective of the present work is to automate the process and achieve a most efficient network resource allocation. To do so, the KQI estimation framework shown in Figure\ref{fig:arq} is proposed for the support of network slice negotiations.
    
    This is composed of four main blocks: Service Experience Acquisition (SEA), Modelling System (ModSys), Operator Slice Negotiation Agent (OSNA) and the Dynamic Slice Allocation (DySA). These blocks have different functionalities as part of two main different stages: the training of the system and its operational phase.
    
    The training phase is dedicated to gather data at different layers and elements of the cellular network and generate the ML models able to estimate the KQIs during the operational phase. In this, service acquisition UEs (SaUEs), this means, UEs where the operator have access to application layer metrics (e.g. drive test terminals, normal users with apps for the indicators extraction, etc.), are required. From this, the main activity of this stage is performed by the Service Experience Acquisition (SEA) and the KQI Modelling system (ModSys). 
    The operational phase is focusing on the support of the slice negotiation and maintenance, using the models created in the training phase.
    
    \vspace{-0.4\baselineskip} 
    
    \subsection{Service Experience Acquisition (SEA)}
    \label{sec:SEA}
    This block is dedicated to acquire the measurements needed from the SaUEs and the network elements for the posterior modelling of the KQIs of the service. To this end, the system tests multiple configurations (e.g. different bandwidth) of the network in different radio conditions (e.g. low and high coverage environments), executing for each of them multiple instances of the service. This process can be automated through the control of the SaUEs and the OAM platform of the network.

    %This block is dedicated to gather the set of measurements or training database required to the posteior modelling of the KQIs of the service. To do so, the system tests multiple configurations of the network, network conditions and base stations, executing for each of them multiple instances of the service. This process can be automated through the control of the OAM platform of the network and the SaUE.
    
    %For the implementation of the SaUE, a computer connected to the network through an LTE stick has been used. This allows for the gathering of the general radio conditions of the terminal: RSRP (\textit{Reference Signal REceived Power}), RSRQ (\textit{Reference Signal Received Quality}), or RSSI (\textit{Received Signal Strength Indicator}).
    
    For video services, the SaUEs executes multiple video playbacks, obtaining their KQIs. These KQIs of the video service as defined by the 3GPPP \cite{3gpp.32.862}: initial time, video bitrate or video stalls, which links with the moments when the image is frozen.    At the same time, the radio conditions of the network are measured by parameters such as RSRP (\textit{Reference Signal Received Power}), RSRQ (\textit{Reference Signal Received Quality}), or RSSI (\textit{Received Signal Strength Indicator}), as well as the network configuration (OAM data) used during the different playbacks.

    %With that objective, the service is launched in first instance with the help of a software framework called Selenium, which lets automation and application testing in any browser. In background, a Python script processes all the data from the DASH client in order to acquire the KQIs of the video service as defined by the 3GPPP \cite{3gpp.32.862}: initial time, video bitrate or video stalls, which links with the moments when the image is frozen.
    
    In order to properly acquire all these data, the SEA block defines and executes a measurement campaign 
    through different calls to the SaUEs, where the duration of the campaign $T$ can be estimated as:
 \vspace{-0.2\baselineskip} 
       	\begin{equation}
    	    \label{eq:durationExp}
            T = \beta \cdot \gamma  \cdot ( n \cdot (\iota + \Delta \iota)+\tau) ,
        \end{equation}
        
     where $\beta$ is the number of base stations where the measurement campaign will be performed, $\gamma$ the amount of possible slice configurations to be tested. The number of service executions, i.e. video playbacks, for each configuration is represented by $n$ where $\iota$ corresponds with the video length and $\Delta \iota$ the time required between executions to relaunch the experiment. Finally, $\tau$ represents the slice reconfiguration time.  
    % In order to reduce the time spent in this phase, design of experiments (DOE) approaches might be required \cite{DOE}.
    
    %In order to properly acquire the desire training database, the block defines and executes a measurement campaign based on the available SaUEs and the OAM or slices possible configurations in terms of amount of resources or parameters, where the  duration, of the campaign can be estimates as:

        \vspace{-0.6\baselineskip} 

    \subsection{Modelling System (ModSys)}
    \label{sec:ModSyS}
    Continuing the training phase, the data gathered by the SEA block is stored in the training database. From this, the ModSys is in charge of generating the KQIs estimation functions from both SaUEs and cellular network data (KPIs and CMs). In this way, for each KQI, $\rho$, a regression function $f$ shall be defined, such as:
    
    % TODO: Nos falta meter el impacto de las condiciones fisicas (RSRP, RSRQ) deberíamos ponerlo como una variable de radio conditions y quizás habría que meter de nuevo lo de el diseño de experimentos
    
    	\begin{equation}
       \varphi'(t)  = f({\boldsymbol\Psi(t)}, {\boldsymbol\vartheta(t)}, {\boldsymbol\Gamma(t)})
 \end{equation}

   where $\varphi'(t)$ denotes the estimated value of $\varphi(t)$ at instant $t$, calculated by the regression function $f$. This function takes as inputs  {$\boldsymbol\Psi(t)$}, which represents the set of measured KPIs, {$\boldsymbol\vartheta(t)$} corresponding to the different CMs of the network and $\boldsymbol\Gamma(t)$ are the radio conditions like RSRP or RSRQ.
   
   The construction of these models can be performed by different regression techniques\cite{LR}. As these techniques can have various accuracy for different KQIs and conditions, for each KQI the different techniques are used to simultaneously train regressions models. These models are then evaluated using k-fold cross-validation with the training data. Their performance is then measured in terms of the coefficient of determination ({$R^2$}) \cite{LR} calculated over the training data.
   
   \iffalse
   by the expression:
       
       \vspace{-0.2\baselineskip} 
    \begin{equation}
    R^{2} = \frac{}{}= \frac{\sum ({\varphi'} - \bar{{\varphi'}})^{2}}{\sum (\varphi - \bar{\varphi})^{2}},
    	\label{eq:rsq}
    \end{equation}

   %_{\forall{\varphi}\in \bold T}
   where $\bar{\varphi}$ and $\bar{\varphi'}$ represents the mean of the estimated and ${\varphi}$ the measured values. This expression can be seen also as the normalization of the standard deviation of the residuals, that is the RMSE (Root Mean Square Error).
    \fi
    
    %For each KQI, those models showing a better accuracy are then selected to be used during the online phase. 
    
    During the operational phase, the estimator that obtained the best performance is used. For each KQI, those models showing a better accuracy are then selected to be used during the online phase. In this phase, ModSys will be called in order to generate KQIs estimation based on the current available low-layer indicators and the possible slices configuration.Additional retraining or online training and best estimator selection can be also implemented whenever relevant new data coming from SaUEs is available.

  \vspace{-0.8\baselineskip} 
    \subsection{Operator Slice Negotiation Agent (OSNA)}
    \label{sec:OSNA}
   At any point during the operational phase, a vertical industry triggers the network slice negotiation with the network operator. The selected estimators from the ModSys block will then be used by the operator slice negotiation agent (OSNA) to support the estimation of the required configuration/resources (and therefore pricing) for the expected radio conditions. This capability would be key to proper implementation of slices negotiated with verticals as well as for the operator to finely tune the resources required to provide the proper performance to their clients.
          \vspace{-0.6\baselineskip} 
    \subsection{Dynamic Slice Allocation (DySA)}
    \label{sec:DySA}
    %From the ModSys block, the selected estimators can be used to support the estimation of pricing during the negotiation phase, the application of the specific slice resources based on the expected radio conditions when this is created. Also the allocation of resources for the slice can dynamically change based on the variable radio conditions (e.g. if the UEs associated to the slice move to an area with poor coverage) in order to maintain the quality of service agreed with the vertical. This capability would be key to proper implementation of slices negotiated with verticals as well as for the operator to finely tuned the resources required to provide the proper performance to their clients.
    Once a slice has been negotiated, the experienced KQIs of the UEs in that slice can dynamically change based on the variable radio conditions (e.g. if the UEs associated to the slice move to an area with poor coverage). Where classical approaches consider static allocation of resources for the slice, the proposed system introduces the concept of Dynamic Slice Allocation (DySA).
    
    In this way, in order to maintain the quality of service agreed with the vertical, the DySA block is in charge of using the ModSys estimators to adapt the radio resources accordingly. To do so, the DySA monitors the low-layer metrics from the network operator OAM and control plane (for the UEs radio conditions). Although it can be done in different ways, in our implementation a RESTful interface is used.
     
     In order to establish the proper slice configuration a set of automatically generated thresholds are compared with the estimated KQIs for each possible resource configuration of the network under and considering the current radio and network conditions. These thresholds are constructed based on the available regression models in the ModSys by:
    \vspace{-0.2\baselineskip} 
    \begin{equation}
	    \label{eq:TH}
      \varrho (t) = \varphi(t) + \alpha 
    \end{equation}

    In this expression, $\alpha$
    represents the security margin, that can be estimated based on the performance of the model during the training phase. $\varphi (t)$ denotes the value of the KQI which is estimated by the ML model selected by the ModSys block. By last, $\varrho(t)$ represents the reached value of the KQI with a network configuration in a time instant $t$.

    In this way, the DySA, by the selection of the configuration whose threshold has a value closer and compliant to the target KQI requisite,  is able to get the adequate configuration  corresponding to the negotiated slice conditions.

      \vspace{-0.4\baselineskip} 
    
	\section{Evaluation}
	\label{sec:evaluation}
	In order to evaluate the estimation capabilities of the proposed system, a wide data-set is built and then applied for the assessment and analysis of key different regression techniques: Linear Regression (LR) \cite{LR}, Stepwise Linear Regression (SW-LR) \cite{LR}, Decision Tree (DTR) \cite{DT}, Gaussian (SVM-G), Cubic (SVM-C) and Quadratic (SVM-Q) Support Vector Machine \cite{SVM} and Gaussian Process Regression (GPR) \cite{GP}.
	
	With the aim to acquire the training data, the video service is launched in a SaUE (based on a computer connected to a LTE testbed \cite{smartCampus} through an LTE stick) with the help of a software framework called Selenium, which lets automation and application testing in any browser. In background, a Python script processes all the data from the DASH client in order to acquire the KQIs of the video service.

    \vspace{-0.7\baselineskip} 

	\subsection{Estimation performance}
    
    % TODO: Falta meter el numero de experimentos, de celdas, de configuraciones, etc.--- DONE
  
\begin{figure}[h] %[btp]
		\centering
    \includegraphics[width=\columnwidth]{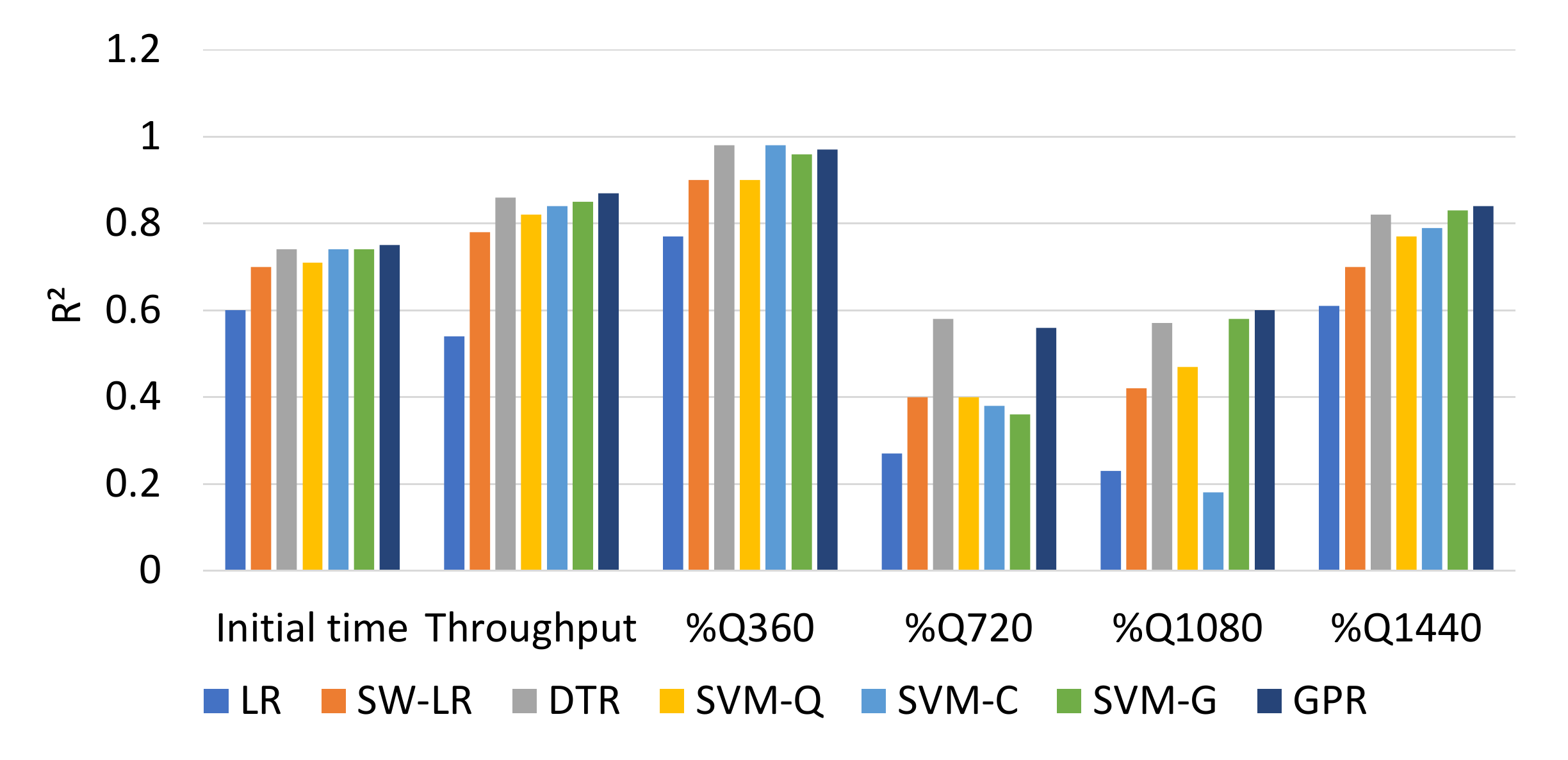}
		\caption{Coefficient of determination ({$R^2$})}
		\label{fig:rsquared}
	\end{figure}

    A dataset of 800 playbacks in 4 different base stations and with 4 different configurations, for   10-fold cross-validation with 70\% of data for training and 30\% for testing is performed. Figure \ref{fig:rsquared} shows the value of the $R^2$ for the different evaluated regression mechanisms presented in subsection \ref{sec:ModSyS} for the different video service KQIs: initial time, average throughput and the percentage of video watched at each quality during the playback \cite{3gpp.32.862}. For the latest we distinguish between the percentage of the video played at each resolution: 360p, 720p, 1080p and 1440p, represented in the figure as \%Q360p, \%Q720p, \%Q1080p and \%Q1440p, respectively.

    As it can be observed in Figure \ref{fig:rsquared}, the models generated by GPR and DTR present the best values of {$R^2$} for all KQIs. SVM, in their three variants, also achieves good estimation values ({$R^2$}{$>$}0.8). Moreover, it is observed that the models for 720p and 1080p resolution have worse performance than the others. This is due to the video client common changes between these qualities, which hinders the estimation. However, with the model obtained for 360p, values of {$R^2$} close to 1 are obtained. In this way, estimation errors can be overcome in order to assure the appropriate performance of the slice configured by the DySA.

	       \vspace{-0.8\baselineskip} 
	 %\vspace{-0.5\baselineskip} 
% TODO: hay que quitar una referencia

\begin{figure}[h] %[btp]
		\centering
		\includegraphics[width=\columnwidth]{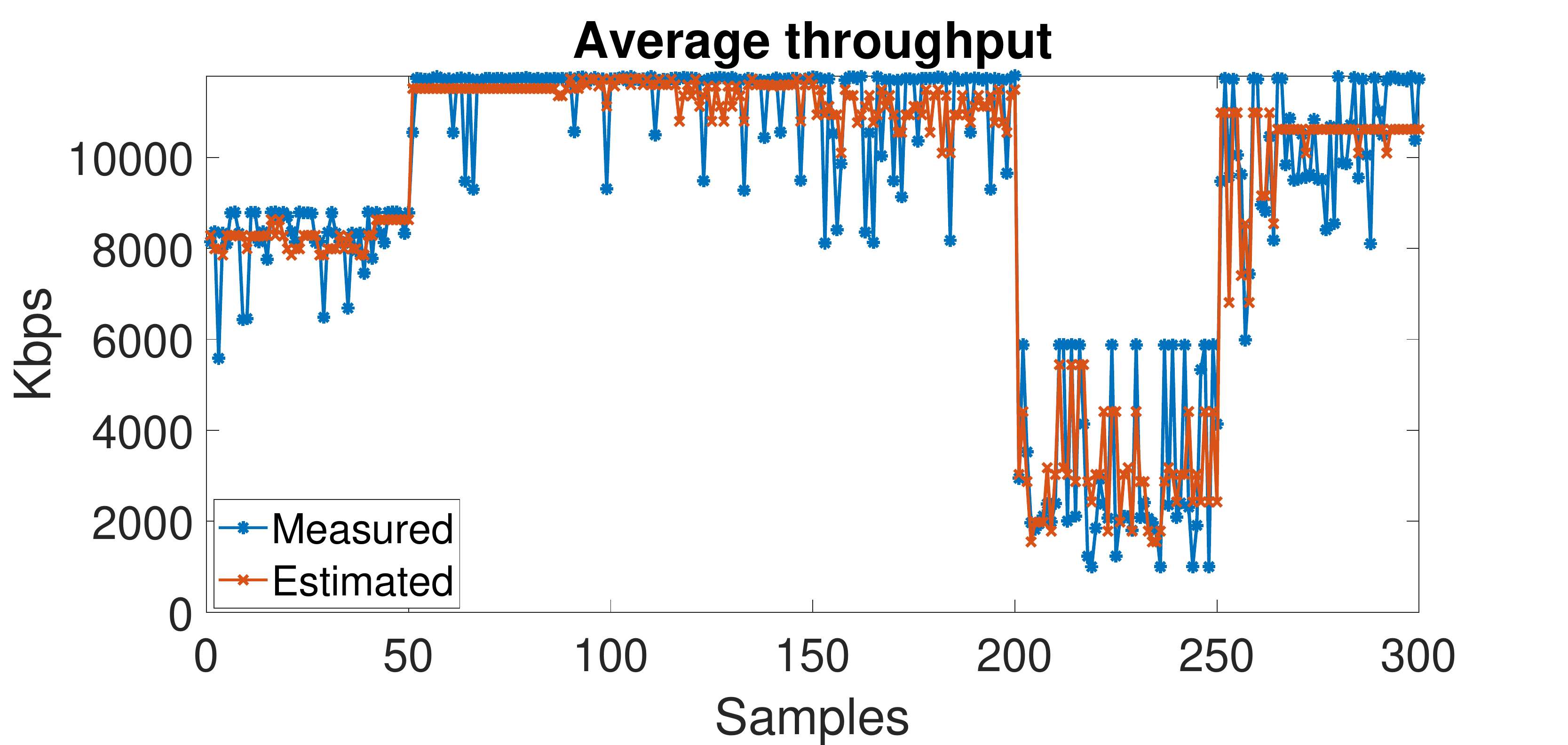}
		\caption{Measured and estimated average throughput}
		\label{fig:estimation}
	\end{figure}

As an example of the system capabilities, Figure \ref{fig:estimation} presents the prediction of the mean throughput of video by the DTR model. As it can be observed, the estimated values are close to the actual measured KQIs, although there are few outliers. Nonetheless, these are given by the dynamic functionality of DASH, which sometimes can change between two consecutive qualities. These outliers are taken in account in the DySA thresholds by the security margins.

    %Gaussian process (GPR), Support Vector Machine Quadratic (SVM Q), Cubic (SVM C) and Gaussian (SVM G), Regression tree (DTR), Stepwise Linear Regression (SW-LR) and Linear Regression (LR). Moreover, it is used cross-validation technique, which consists in the creation of different subsets to train and test the model being in this case 70\% for training and 30\% for testing, as well as it is used a 10-fold cross-validation. As it can be seen, GP and DTR present the best values of {$R^2$} for all KQIs, as well as SVM, in their three variants, gets good values for this coefficient too.

	%In this way,  7 different regression techniques are evaluated: linear regression (LR), stepwise linear regression (SW-LR), decision tree (Fine tree), support vectorial machine quadratic (SVM-Q), cubic (SVM-C) and gaussian (SVM-G), as well as gaussian regression process (Gaussian-rational). As it can be seen in figure \ref{fig:rsquared}, both fine tree and gaussian-rational show a good estimations of the indicators.

	Additionally, the estimation time of the models, this is, how long it takes for them to calculate the KQI' from their inputs is also measured as it is key for how fast the OSNA and the DySA can obtain new calculation of resources. The training time is not so critical as the training is to be performed just for the generation of the models and sparsely for their retraining.

		\begin{figure}[h] %[btp]
		\centering
		\includegraphics[width=0.75\columnwidth]{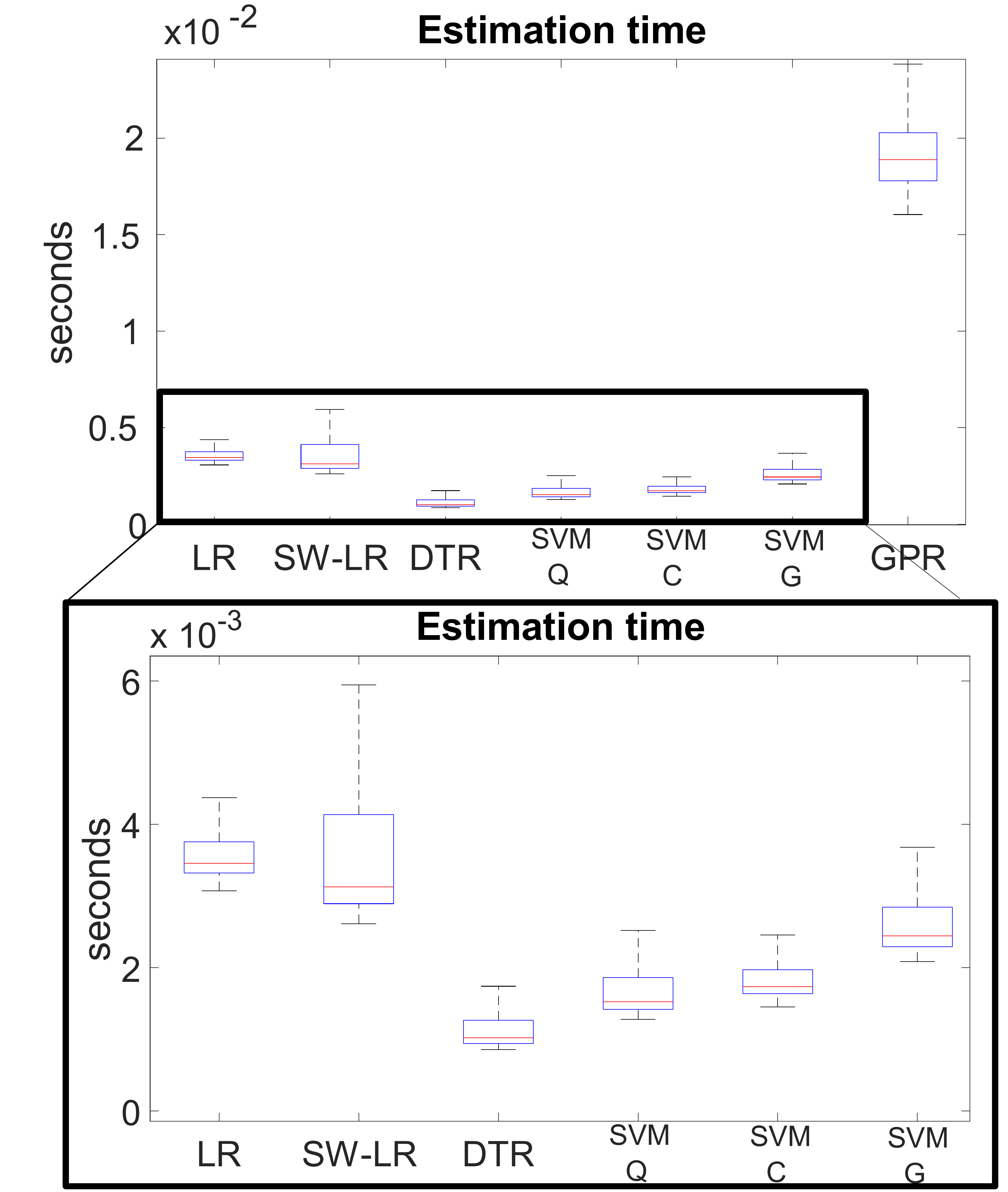}
		\caption{Estimation time of different machine learning mechanisms}
		\label{fig:time}
	\end{figure}
	 
	Hence, this distribution of the estimation times for 1000 executions of the algorithms for the average throughput in a common PC, with an Intel Core i7-8550u, are represented in the boxplots of figure \ref{fig:time}. As it can be seen, DTR is not only one of the best techniques in terms of performance, but also it is the fastest technique. Nevertheless, GPR, being the other technique that better estimates the KQIs with a slightly superior accuracy, requires far more time than the other mechanisms due to the higher complexity of the regression function defined by this model \cite{GP}.
	
	%, where first, second and third quartiles are used to form the box.
	
	% TODO: "due to the higher complexity in model evaluation." No lo entiendo

	\section{Conclusions}
	\label{sec:conclusion}
	Although network slice negotiation processes are expected to be one of the main characteristics of 5G networks, the translation from the high-layer E2E requirements of users and verticals to specific radio-access slice configurations has been mainly not addressed yet. In this area, this work has presented a framework for the application of KQI estimation to support the slice negotiation, allocation of resources and dynamic maintenance, with a special focus on video streaming services.
	The defined system has been evaluated in a real indoor cellular network and for realistic streaming conditions and protocols. Results have shown that the proposed ML algorithms provide reliable estimation of the quality perceived by users. %This paves the way to network management functions for obtaining and maintaining required users' experience via an accurate and dynamic allocation of slice radio resources.

	%This paves the way, network management activities can be oriented towards reaching and maintain the quality of the user's experience and enabling an accurate and dynamic allocation of slice radio resources. 

%--------------------------------------------------------------------------------------

	%TC:ignore 
	%TC:endignore 

%\bibliographystyle{ieeetr}
	
\hypersetup{
    colorlinks=true,
    citecolor=black,
    linkcolor=black,     
    urlcolor=black,
}
	\bibliographystyle{IEEEtran}
	\bibliography{Bibliography} 

%	\end{thebibliography}
\end{document}